\documentclass[longauth]{aa} 
\usepackage{graphicx,natbib}
\usepackage{txfonts}
\begin{document}
   \title{Radio-to-UV monitoring of \object{AO 0235+164} by the WEBT and Swift 
   during the 2006--2007 outburst\thanks{ The radio-to-optical data 
   presented in this paper are stored in the WEBT archive; for questions regarding their availability,
   please contact the WEBT President Massimo Villata ({\tt villata@oato.inaf.it}).}}


   \author{C.~M.~Raiteri              \inst{ 1}
   \and   M.~Villata                  \inst{ 1}
   \and   V.~M.~Larionov              \inst{ 2,3}
   \and   M.~F.~Aller                 \inst{ 4}
   \and   U.~Bach                     \inst{ 5}
   \and   M.~Gurwell                  \inst{ 6}
   \and   O.~M.~Kurtanidze            \inst{ 7}
   \and   A.~L\"ahteenm\"aki          \inst{ 8}
   \and   K.~Nilsson                  \inst{ 9}
   \and   A.~Volvach                  \inst{10}
   \and   H.~D.~Aller                 \inst{ 4}
   \and   A.~A.~Arkharov              \inst{ 3}
   \and   R.~Bachev                   \inst{11}
   \and   A.~Berdyugin                \inst{ 9}
   \and   M.~B\"ottcher               \inst{12}
   \and   C.~S.~Buemi                 \inst{13}
   \and   P.~Calcidese                \inst{14}
   \and   E.~Cozzi                    \inst{15}
   \and   A.~Di Paola                 \inst{16}
   \and   M.~Dolci                    \inst{17}
   \and   J.~H.~Fan                   \inst{18}
   \and   E.~Forn\'e                  \inst{19}
   \and   L.~Foschini                 \inst{20}
   \and   A.~C.~Gupta                 \inst{21}
   \and   V.A.~Hagen-Thorn            \inst{ 2}
   \and   L.~Hooks                    \inst{12}
   \and   T.~Hovatta                  \inst{ 8}
   \and   M.~Joshi                    \inst{12}
   \and   M.~Kadler                   \inst{22}
   \and   T.~S.~Konstantinova         \inst{ 2}
   \and   A.~Kostov                   \inst{23}
   \and   T.~P.~Krichbaum             \inst{24}
   \and   L.~Lanteri                  \inst{ 1}
   \and   L.~V.~Larionova             \inst{ 2}
   \and   C.-U.~Lee                   \inst{25}
   \and   P.~Leto                     \inst{26}
   \and   E.~Lindfors                 \inst{ 9}
   \and   F.~Montagni                 \inst{27}
   \and   R.~Nesci                    \inst{28}
   \and   E.~Nieppola                 \inst{ 8}
   \and   M.~G.~Nikolashvili          \inst{ 7}
   \and   J.~Ohlert                   \inst{29}
   \and   A.~Oksanen                  \inst{30}
   \and   E.~Ovcharov                 \inst{31}
   \and   P.~P\"a\"akk\"onen          \inst{32}
   \and   M.~Pasanen                  \inst{ 9}
   \and   T.~Pursimo                  \inst{33}
   \and   J.~A.~Ros                   \inst{19}
   \and   E.~Semkov                   \inst{11}
   \and   R.~L.~Smart                 \inst{ 1}
   \and   A.~Strigachev               \inst{11}
   \and   L.~O.~Takalo                \inst{ 9}
   \and   K.~Torii                    \inst{34}
   \and   I.~Torniainen               \inst{ 8}
   \and   M.~Tornikoski               \inst{ 8}
   \and   C.~Trigilio                 \inst{13}
   \and   H.~Tsunemi                  \inst{34}
   \and   G.~Umana                    \inst{13}
   \and   A.~Valcheva                 \inst{11,31}
 }

   \offprints{C.~M.~Raiteri}

   \institute{
          INAF, Osservatorio Astronomico di Torino, Italy                                                     
   \and   Astron.\ Inst., St.-Petersburg State Univ., Russia                                                  
   \and   Pulkovo Observatory, St.\ Petersburg, Russia                                                        
   \and   Department of Astronomy, University of Michigan, MI, USA                                            
   \and   Max-Planck-Institut f\"ur Radioastronomie, Bonn, Germany                                            
   \and   Harvard-Smithsonian Center for Astroph., Cambridge, MA, USA                                         
   \and   Abastumani Astrophysical Observatory, Georgia                                                       
   \and   Mets\"ahovi Radio Obs., Helsinki Univ.\ of Technology, Finland                                      
   \and   Tuorla Observatory, Univ.\ of Turku, Piikki\"{o}, Finland                                           
   \and   Radio Astron.\ Lab.\ of Crimean Astroph.\ Observatory, Ukraine                                      
   \and   Inst.\ of Astron., Bulgarian Acad.\ of Sciences, Sofia, Bulgaria                                    
   \and   Department of Physics and Astronomy, Ohio Univ., OH, USA                                            
   \and   INAF, Osservatorio Astrofisico di Catania, Italy                                                    
   \and   Oss.\ Astronomico della Regione Autonoma Valle d'Aosta, Italy                                       
   \and   New Millennium Observatory, Italy                                                                   
   \and   INAF, Osservatorio Astronomico di Roma, Italy                                                       
   \and   INAF, Osservatorio Astronomico di Collurania Teramo, Italy                                          
   \and   Center for Astrophysics, Guangzhou University, China                                                
   \and   Agrupaci\'o Astron\`omica de Sabadell, Spain                                                        
   \and   INAF, IASF-Bologna, Italy                                                                           
   \and   YNAO, Chinese Academy of Sciences, Kunming, China                                                   
   \and   NASA/Goddard Space Flight Center, Greenbelt, Maryland, USA                                          
   \and   Inst.\ of Astronomy, BAN, Bulgaria                                                                  
   \and   Max-Planck-Institut f\"ur Radioastronomie, 53121 Bonn, Germany                                      
   \and   Korea Astronomy and Space Science Institute, South Korea                                            
   \and   INAF, Istituto di Radioastronomia, Sezione di Noto, Italy                                           
   \and   Greve Observatory, Italy                                                                            
   \and   Dept.\ of Phys., ``La Sapienza" Univ., Roma, Italy                                                  
   \and   Michael Adrian Observatory, Trebur, Germany                                                         
   \and   Hankasalmi Observatory, Finland                                                                     
   \and   Sofia University, Bulgaria                                                                          
   \and   Jakokoski Observatory, Univ.\ of Joensuu, Finland                                                   
   \and   Nordic Optical Telescope, Santa Cruz de La Palma, Spain                                             
   \and   Osaka University, Japan                                                                             
 }

   \date{}

 
  \abstract
   {The blazar \object{AO 0235+164} was claimed to show a quasi-periodic behaviour in the radio and optical 
bands in the past, with the main outbursts repeating every 5--6 years. 
However, the predicted 2004 outburst did not occur, and further analysis suggested a longer time scale, according to which the next event would have occurred in the 2006--2007 observing season. 
Moreover, an extra emission component contributing to the UV and soft X-ray flux was detected, whose nature is not yet clear.
An optical outburst was observed in late 2006 -- early 2007, 
which triggered a Whole Earth Blazar Telescope (WEBT) campaign as well as target of opportunity (ToO) observations by the Swift satellite.}
   {In this paper, we present the radio-to-optical data taken by the WEBT together 
with the UV data acquired by the UltraViolet and Optical Telescope (UVOT) instrument onboard Swift to investigate both the outburst behaviour at different wavelengths and the nature of the extra emission component.}
   {Multifrequency light curves have been assembled with data from 27 observatories; 
optical and UV fluxes have been cleaned from the contamination of the southern active galactic nucleus (AGN). 
We have analysed spectral energy distributions at different epochs, corresponding to different brightness states; extra absorption by the foreground galaxy has been taken into account.}
   {We found the optical outburst to be as strong as the big outbursts of the past: starting from late 
September 2006, a brightness increase of $\sim 5$ mag led to the outburst peak in February 19--21, 2007. 
We also observed an outburst at mm and then at cm wavelengths, with an increasing time delay going toward lower frequencies during the rising phase. 
Cross-correlation analysis indicates that the 1 mm and 37 GHz flux variations lagged behind the $R$-band ones by about 3 weeks and 2 months, respectively. These short time delays suggest that the corresponding jet emitting regions are only slightly separated and/or misaligned.
In contrast, during the outburst decreasing phase the flux faded contemporaneously at all cm wavelengths. 
This abrupt change in the emission behaviour may suggest the presence of some ``shutdown" mechanism of intrinsic or geometric nature.
The behaviour of the UV flux closely follows the optical and near-IR one. 
By separating the synchrotron and extra component contributions to the UV flux, we found that they correlate, which suggests that the two emissions have a common origin.}
  {}



   \maketitle
%

\section{Introduction}

\object{AO 0235+164} at redshift $z=0.94$ is one of the best-studied BL Lac objects.
The analysis of its radio and optical light curves extending over 25 years led \citet{rai01} 
to suggest a quasi-periodic occurrence of the main outbursts every $5.7 \pm 0.5$ years. 
The next outburst was predicted around February--March 2004, and a multiwavelength campaign was organised by the Whole Earth Blazar Telescope (WEBT)\footnote{{\tt http://www.to.astro.it/blazars/webt/} \\ \citep[see e.g.][]{vil06,vil07,boe07,rai07b}} to follow the expected event. 
The radio-to-optical observations by the WEBT were complemented by the optical--UV and X-ray data acquired by the XMM-Newton satellite during 3 pointings in January and August 2004, and January 2005, and by optical spectroscopic observations with the 3.6 m Telescopio Nazionale Galileo (TNG). 
The results of this intense observing effort were published by (\citealt{rai05,rai06b,rai06a,rai07a}; see also \citealt{hag07b} for an analysis of colour variability). 
The predicted outburst was not observed, and time-series analysis on the light curves extended to 2005 revealed a possible characteristic variability time scale of $\sim 8$ years.
Moreover, the XMM-Newton observations suggested the presence of an extra emission component in the source spectral energy distribution (SED), in addition to the synchrotron and inverse-Compton ones.
The origin of this component, peaking in the UV/soft X-ray frequency range, was ascribed either to thermal emission from an accretion disc, or to synchrotron emission from an inner jet region. 

An increased radio activity was detected in the 2005--2006 observing season \citep{bac07}, and a
dramatic optical brightening was finally observed in late 2006 -- early 2007. 
This triggered a new WEBT multiwavelength campaign as well as ToO observations by the Swift satellite.
In this paper, we present the radio-to-optical observations performed from spring 2005 (i.e.\ the end of the period considered
in \citealt{rai06b}) to October 2007, 
and the data acquired by the UltraViolet and Optical Telescope (UVOT) instrument onboard Swift. 
X-ray data from the Swift X-ray Telescope (XRT) and Burst Alert Telescope (BAT) instruments will be presented in another paper (Kadler et al., in preparation).

This paper is organised as follows:
the procedures we adopted to treat the WEBT and UVOT data are outlined in Sect.~2, with particular attention paid to the subtraction of the southern AGN contribution from the optical and UV photometry.
The multifrequency light curves are presented and discussed in Sect.~3 while in Sect.~4 time lags among variations at different frequencies are derived by means of statistical analysis.
Spectral energy distributions with simultaneous data from near-IR to UV are constructed in Sect.~5, with the aim of separating the synchrotron and the extra emission components and of understanding their relationship. Finally, the results of our work are discussed in Sect.~6. 

\section{Data reduction and analysis}

\subsection{Radio-to-optical observations}

   \begin{figure*}
   \sidecaption
   \includegraphics[width=12cm]{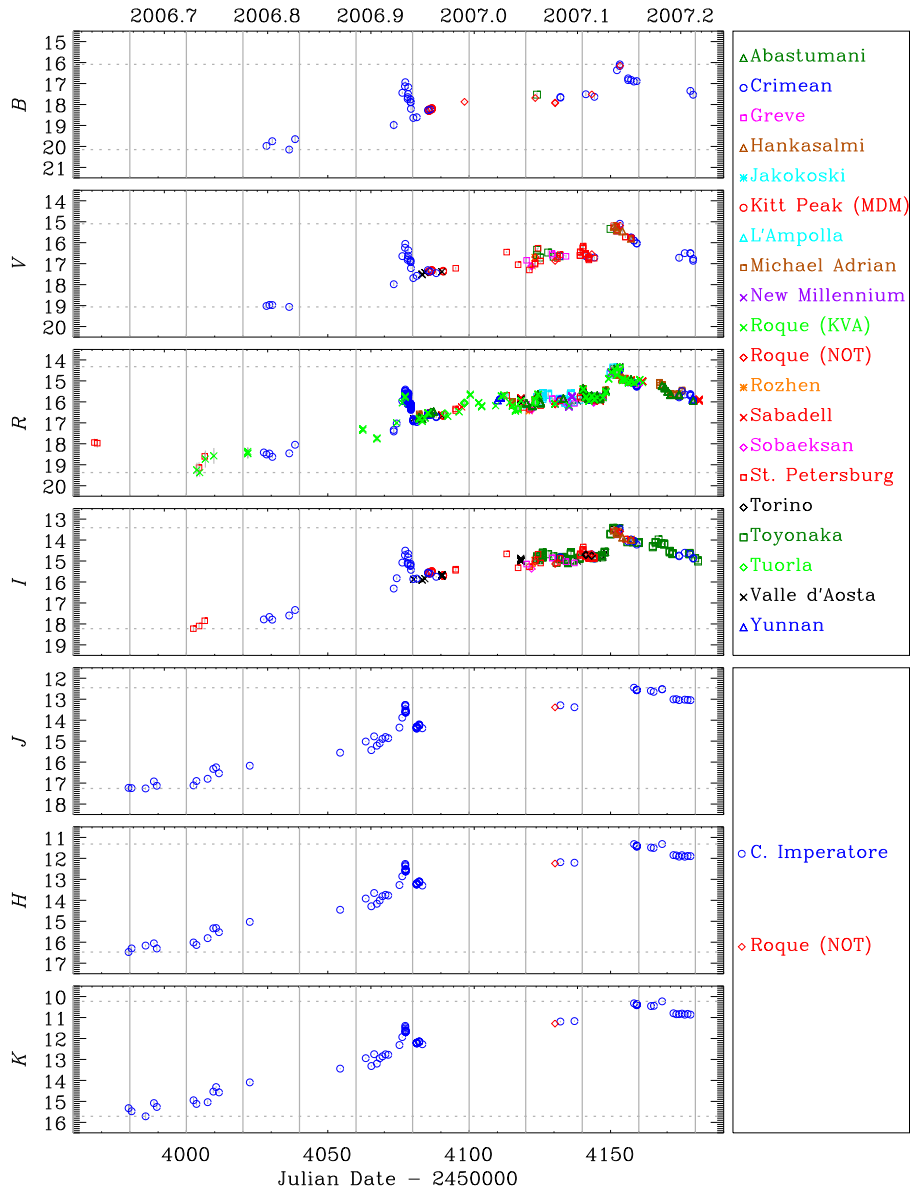}
      \caption{Optical and near-IR light curves of AO 0235+164 in the 2006--2007 observing season. 
The optical magnitudes are corrected for the ELISA contribution. 
Maximum and minimum brightness levels are highlighted with horizontal dotted lines.}
         \label{ottico-nir}
   \end{figure*}

The new WEBT campaign on AO 0235+164 saw a wide international participation.
Data were collected starting from the end of the period considered in \citet{rai06b}, i.e.\ from spring 2005 to October 2007.
Table \ref{obs} shows the list of participating observatories, the size of the telescopes,
and the bands/frequencies at which we acquired the data.

We collected optical and near-IR data as instrumental magnitudes of the source and comparison stars 
in the same field. 
Magnitude calibration was performed using preferably Stars 1, 2, and 3 by \citet{smi85} in the optical bands\footnote{In some cases, we also used Stars 9, 10, and 11 by \citet{gon01}.}, and the \citet{gon01} photometric sequence in the near-IR. 
Optical light curves were carefully assembled and cleaned \citep[see e.g.][]{vil02}.

We collected radio data as calibrated flux densities; no systematic offset among datasets provided by different observers was evident; 
data binning was needed in a few cases to reduce the noise. 
We also included data from the VLA/VLBA Polarization Calibration Database\footnote{\tt http://www.vla.nrao.edu/astro/calib/polar/} 
(PCD) in our light curves.

\begin{table}
\caption{Ground-based observatories participating in this work.}
\label{obs}
\centering
\begin{tabular}{l r c  }
\hline\hline
Observatory    & Tel.\ size     & Bands\\
\hline
\multicolumn{3}{c}{\it Radio}\\
Crimean (RT-22), Ukraine & 22 m          & 22, 37 GHz          \\
Mauna Kea (SMA), USA     &$8 \times 6$ m$^a$ & 0.85, 1 mm      \\
Medicina, Italy          & 32 m          & 5, 8, 22 GHz        \\
Mets\"ahovi, Finland     & 14 m          & 37 GHz              \\
Noto, Italy              & 32 m          & 43 GHz              \\
UMRAO, USA               & 26 m          & 5, 8, 14.5 GHz      \\
\hline
\multicolumn{3}{c}{\it Near-infrared}\\
Campo Imperatore, Italy  & 110 cm        & $J, H, K$          \\
Roque (NOT), Spain       & 256 cm        & $J, H, K$          \\
\hline
\multicolumn{3}{c}{\it Optical}\\
Abastumani, Georgia      &  70 cm        & $R$                \\
Crimean, Ukraine         &  70 cm        & $B, V, R, I$       \\
Greve, Italy             &  32 cm        & $R$                \\
Hankasalmi, Finland      &  40 cm        & $V, R, I$          \\
Jakokoski, Finland       &  50 cm        & $R$                \\
Kitt Peak (MDM), USA     & 130 cm        & $B, V, R, I$       \\
L'Ampolla, Spain         &  36 cm        & $R$                \\
Michael Adrian, Germany  & 120 cm        & $R$                \\
New Millennium, Italy    &  36 cm        & $R$                \\
Roque (KVA), Spain       &  35 cm        & $R$                \\
Roque (NOT), Spain       & 256 cm        & $U, B, V, R, I$    \\
Rozhen, Bulgaria         & 50/70 cm      & $B, V, R, I$       \\
Sabadell, Spain          &  50 cm        & $R$                \\
Sobaeksan, South Korea   &  61 cm        & $V, R, I$          \\
St.\ Petersburg, Russia  &  40 cm        & $V, R, I$          \\
Torino, Italy            & 105 cm        & $R, I$             \\
Toyonaka, Japan          &  35 cm        & $B, V, R, I$       \\
Tuorla, Finland          & 103 cm        & $R$                \\
Valle d'Aosta, Italy     &  81 cm        & $V, R, I$          \\
Yunnan, China            & 102 cm        & $R$                \\
\hline
\multicolumn{3}{l}{$^a$ Radio interferometer including 8 dishes of 6 m size.}
\end{tabular}
\end{table}

\subsection{UV observations by Swift}

The Swift satellite observed the source from January 28 to February 26, 2007.
The UVOT instrument \citep{rom05} acquired data in the $U$, UV$W1$, UV$M2$, and UV$W2$ filters, which
were processed with the {\tt uvotmaghist} task of the HEASOFT 6.3 package.
Following the recommendations contained in the release notes, we used an aperture radius of 5 arcsec
in agreement with the standard photometric aperture defined in the calibration files (CALDB updated as July 2007). 
Background was extracted in an annulus centred on the source, with inner and outer radii of 8 and 18 arcsec, respectively. 

\subsection{Correction for ELISA contribution}

A major point in the analysis of the optical and UV radiation coming from AO 0235+164 is to eliminate the contamination from the southern AGN, named ELISA by \citet{rai05}. 
Since it lies only 2 arcsec south of the source and it is very faint \citep{nil96}, this object is not resolved from AO 0235+164 in most optical frames as well as in the UVOT ones.
Consequently, it gives a contribution to the AO 0235+164 photometry, which affects more the blue than the red part of the optical--UV spectrum and is stronger when the blazar is fainter.
To correct for this effect, \citet{rai05} estimated the $UBVRI$ magnitudes of ELISA from images taken at medium-size telescopes participating in the 2003--2004 WEBT campaign.
More recently, \citet{rai07a} analysed an ELISA optical spectrum taken at the 8 m Very Large Telescope (VLT). The $UBVRI$ magnitudes derived from this spectrum confirm the 
previous photometry results and lower the uncertainties to 0.05 mag; only in the case of the $V$ filter a slightly fainter value appears more appropriate.
Hence, to subtract the ELISA contribution from the optical fluxes,
we adopted the following ELISA magnitudes:
$U=20.80$, $B=21.40$, $V=20.95$, $R=20.50$, and $I=19.90$.
Zero-magnitude fluxes were taken from \citet{bes98}.

We then estimated the ELISA contribution in the UV$W1$, UV$M2$, and UV$W2$ filters from the HST/FOS composite spectrum published by \citet{bur96}\footnote{We note that the flux in the UV$W1$ band is affected by the contribution of the \ion{C}{iii}] emission line, and that the HST/FOS UV spectrum nicely reconnects in flux with the VLT optical spectrum.}.
We subtracted these ELISA flux densities from the total flux densities obtained with 
{\tt uvotmaghist} and converted the results into magnitudes with zero-magnitude fluxes of
$3.963, \, 4.544, \, 5.400 \, \times 10^{-9} \rm \, erg \, cm^{-2} \, s^{-1} \, \AA^{-1}$. 
UV magnitudes of ELISA are: 20.45, 20.50, and 20.65 in the UV$W1$, UV$M2$, and UV$W2$ bands, 
respectively, again with a 0.05 mag uncertainty.

As in the previous works, the ELISA contamination in the near-IR is assumed to be negligible, 
and null in the mm and cm bands.

   \begin{figure}
   \resizebox{\hsize}{!}{\includegraphics{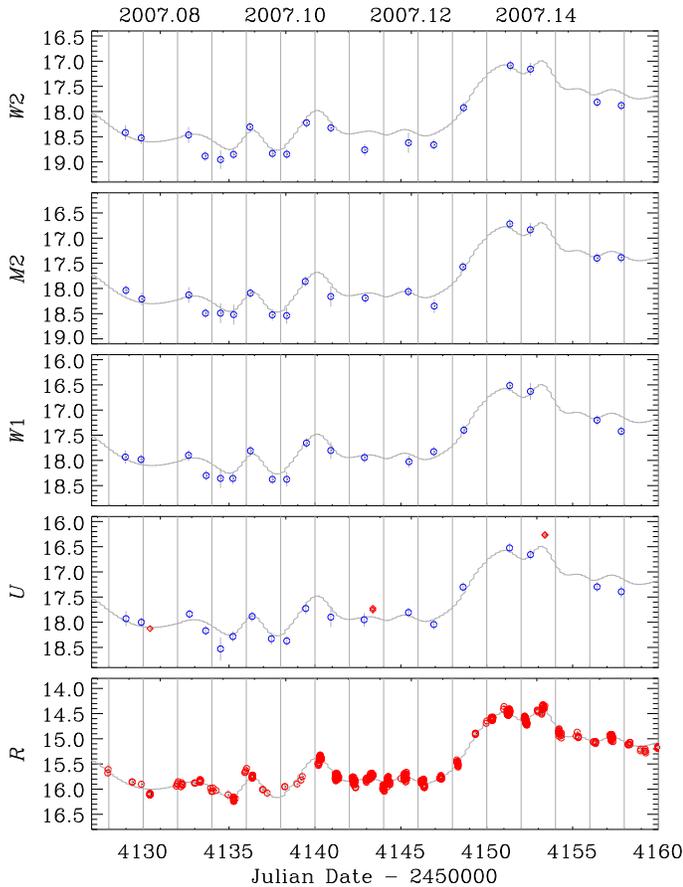}}
      \caption{UVOT light curves of AO 0235+164 compared to the ground-based $U$ data from the 
NOT (red diamonds) and to the $R$-band data (bottom panel). 
All data have been corrected for the ELISA contribution.
The cubic spline interpolation through the 1-day binned $R$-band light curve is shown in each panel, 
properly rescaled (see text).}
         \label{uvot_clean}
   \end{figure}

\section{Light curves behaviour}

\begin{figure*}
   \sidecaption
   \includegraphics[width=12cm]{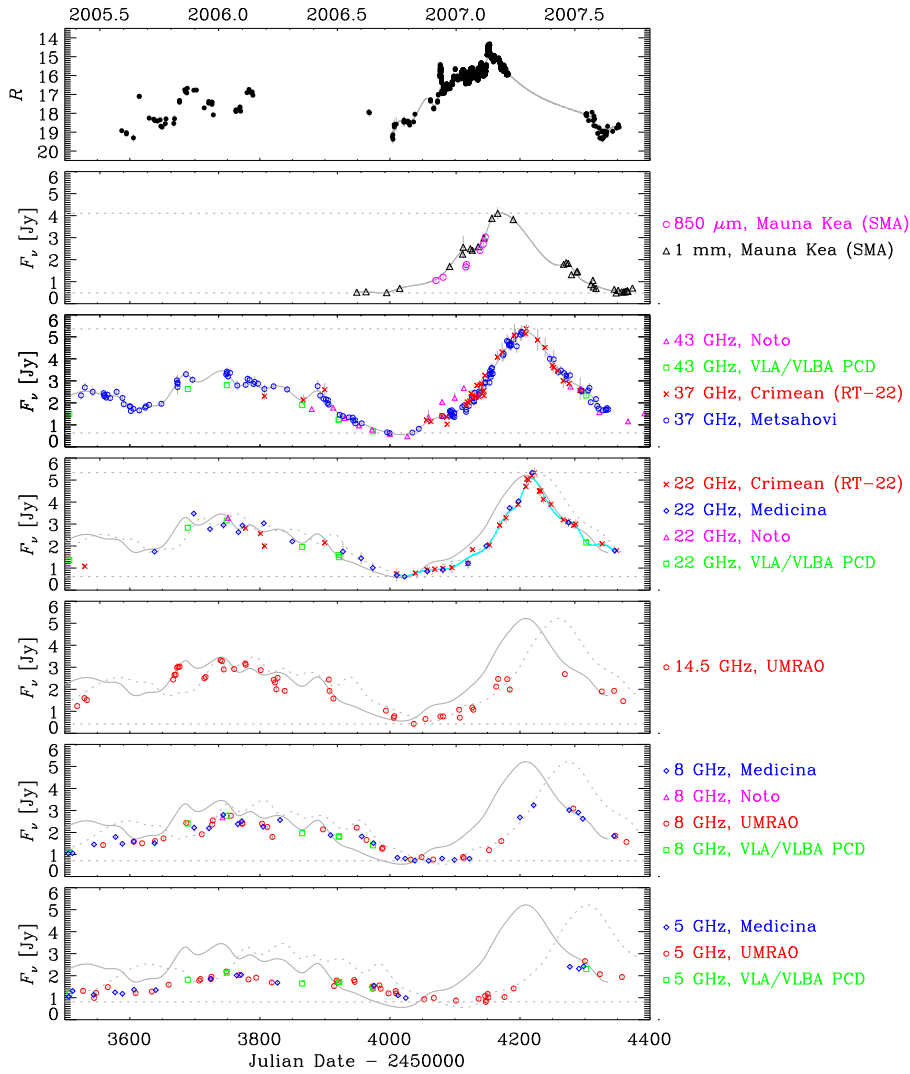}
      \caption{Comparison among the $R$-band magnitudes ({\it top panel}), 
corrected for the ELISA contribution,
and the radio flux densities (Jy) at different frequencies. 
Cubic spline interpolations through the 10-day binned light 
curves in the $R$ and 1 mm bands have been drawn in the first and second panel, respectively.
Similarly, the cubic spline interpolation through the 20-day binned light curve at 37 GHz has been drawn in the other panels, to highlight the time delays between the flux variations at 37 GHz and those at lower frequencies. In the fourth panel the cubic spline interpolation through the 15-day binned light curve at 22 GHz is also shown (cyan line).
Minimum (and maximum) flux density levels are indicated with horizontal dotted lines.}
         \label{radop}
   \end{figure*}

The final light curves are shown in Figs.\ \ref{ottico-nir}--\ref{radop}.
The optical and UV magnitudes have been corrected for the ELISA contribution,
as explained in the previous section. 

Figure \ref{ottico-nir} displays the optical and near-IR light curves of the 2006--2007 observing season.
Notwithstanding the differences in time sampling, the general trend is similar at all wavelengths, as expected.
At the beginning of the observing season the source was in a faint state, but
by in late September the source brightness started to increase.
We observed a fast flare to peak at JD = 2454077.4 (December 7, 2006). 
It was characterized by a 2.0 mag brightening in 4 days in the $R$ band, and
by noticeable intranight variability \citep[see also][]{hag07a}. After the drop of 1.5 mag in 3 days,
the source brightness continued to rise smoothly, with minor flares superposed.
The outburst culminated on JD = 2454151--3 (February 19--21, 2007). This peak is well sampled in the
optical bands, but was missed by the near-IR observations, which registered the brightest level some days later, 
when the optical flux was already in a decreasing phase.
We notice that in this observing season the variability amplitude in the best-sampled $R$ band was $\sim 5$ mag.

Observations by the Swift satellite were triggered in late January during the culminating phase of the outburst.
The $U$ and UV light curves from UVOT are shown in Fig.\ \ref{uvot_clean} as blue circles, and compared to the $R$-band light curve. In the $U$-band panel, ground-based data from the Nordic Optical Telescope (NOT) were plotted as red diamonds. UVOT data have been binned daily, and a lower limit to the error of 0.1 mag has been adopted to take into account statistical and systematic errors.
The grey line in each panel represents a cubic spline interpolation through the 1-day binned $R$-band light curve 
(rescaled according to average colour indices: $U-R = 2.1$; $W1-R = 2.1$; $M2-R = 2.3$; and $W2-R = 2.7$).
It helps to recognize that the trend of the $R$-band light curve is well followed by all the UVOT light curves.
Also, the variability amplitude appears nearly the same in the optical and UVOT bands.

The increase of the optical and near-IR fluxes was followed soon after by an increase of the millimetric and then of the centimetric fluxes, with a progressive time delay going toward longer wavelengths.
Figure \ref{radop} shows the radio light curves starting from the end of the observing period presented in \citet{rai06b}, compared to the $R$-band one (top panel). 
Optical observations in the 2005--2006 season were rather sparse;
they show moderate source activity at a level similar to that observed in the previous season \citep{rai06b}.
The figure also shows that, after the solar conjunction in 2007, the source was again in a very faint optical state.

The second panel from the top displays 1 mm and 850 $\mu$m data from the SubMillimeter Array (SMA).
A cubic spline interpolation through the 10-day binned 1 mm light curve is plotted to better follow the behaviour of the source flux at this wavelength.
The 1 mm flux density increased by a factor $\sim 8$ from the beginning of the 2006--2007 season to the time of the observed maximum, on March 6, 2007. After this date, the only data point
acquired before the solar conjunction epoch suggests that the outburst peak was already over.
Moreover, data after solar conjunction show that the mm flux density had reduced by more than a factor of 2.
We notice that a first hump in the 1 mm light curve, around JD = 2454100, may be correlated to the fast optical flare observed at JD $\sim$ 2454080.

In the following panels, radio light curves at lower and lower frequencies are displayed.
We have plotted the cubic spline interpolation through the 20-day binned light curve at 37 GHz, the best-sampled one, to allow an easier comparison of the flux density behaviour at different wavelengths.
In general, the radio spectrum appears ``inverted": the flux density is higher at higher frequencies. This is not true in the period (JD $\sim$ 2453900--2454050) leading to the minimum flux levels, where the flux density appears higher at longer wavelengths, and during the outburst decreasing phase, where the radio spectrum seems to be flat.

The radio emission had already undergone a rather bright phase in 2005--2006, a kind of ``minor outburst" like those often observed in the past, which sometimes are accompanied by noticeable optical events and sometimes not.
A flux minimum was reached around JD = 2454000--2454100, and then the flux started to increase again, all these variations occurring first at the higher and then at the lower frequencies.
However, the 22 GHz flux density reached a maximum value similar to that observed at 37 GHz almost at the same time, and the flux decrease was also similar and contemporaneous. This is highlighted in the fourth panel of Fig.\ \ref{radop} by the overlapping of the spline through the 37 GHz data with that through the 22 GHz ones.
The lower-frequency light curves are more sparsely sampled, but they too suggest a progressively delayed rising phase, and a contemporaneous decreasing phase.

While the general multifrequency behaviour during the outburst seems in agreement with the predictions of shock-in-jet models \citep[see e.g.][]{hug89,val92}, the decay stage might suggest a more complex picture.
Indeed, it seems as if a kind of ``shutdown" occurred when the perturbation that was responsible for the flux increase reached the jet region emitting at cm wavelengths. The shape of the 37 GHz outburst, sharp and without a plateau, may indicate that the outburst observed at this frequency has already been affected by the shutdown.
The signature of the shutdown would be mainly visible in the 22 GHz light curve, which is characterized by an abrupt flux decay.

\begin{figure*}
   \sidecaption
   \includegraphics[width=12cm]{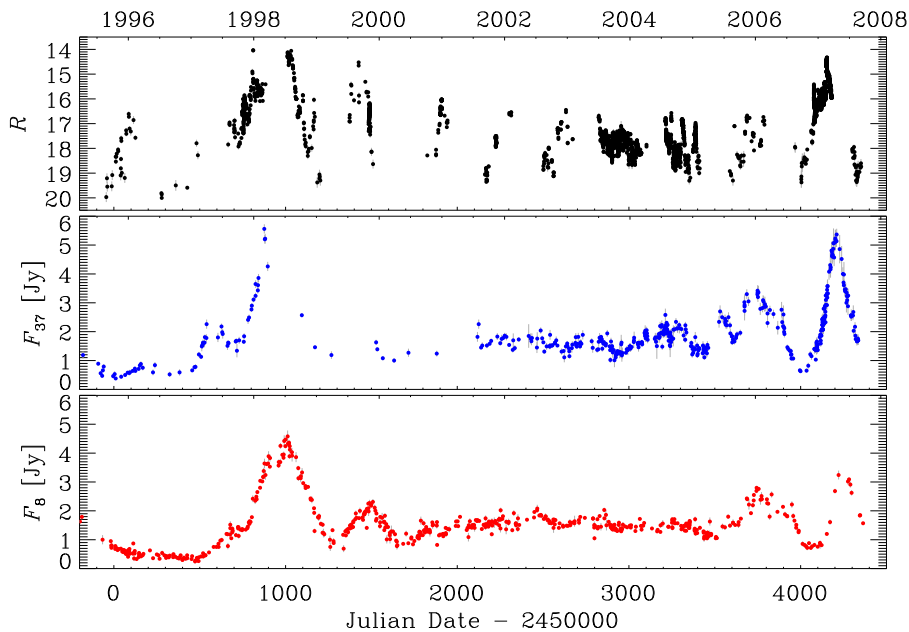}
      \caption{Comparison among the $R$-band magnitudes ({\it top panel}), 
corrected for the ELISA contribution,
and the radio flux densities (Jy) at 37 and 8 GHz over the last $\sim 12.5$ years.}
         \label{long_term}
   \end{figure*}

It is interesting to compare this recent behaviour of the source with its past history.
Figure \ref{long_term} displays the light curves in the optical $R$ band and at two radio frequencies in the last $\sim 12.5$ years, including the previous radio-optical outburst observed in 1997--1998. We notice that the last outburst had a shorter duration with respect to the outburst in 1997--1998. Moreover, if we compare the 2005--2006 and 2006--2007 events, we see that the latter was much ``harder" (i.e.\ stronger at higher frequencies) than the former. Similarly, the 1999--2000 post-outburst event is harder than the preceding major one. Other very hard (optical only) events are visible along all the period shown in Fig.\ \ref{long_term}, starting from the 1995--1996 event \citep{tak98}.
As in the case of 3C 454.3 discussed by \citet{vil07}, these occurrences might be due to (temporary) different orientations of the optical and radio emitting regions, harder events being observed when the optical emitting region is more aligned with the line of sight than the radio one, and hence the corresponding radiation is more beamed.
Alternatively, harder and softer events might be intrinsically different. In this case harder events would correspond to situations where the perturbation vanishes earlier in the jet, sometimes being unable to reach the radio emitting regions.
Regardless of their origin, hard flares in average should be shorter, since they cannot fully develop.

In this framework we could also explain the sharpness of the last radio outburst, and the fast contemporaneous flux decrease at all wavelengths. They might be due either to a geometric effect, i.e.\ a jet bending in the radio emitting region (as proposed for 3C 454.3; see \citealt{vil07}), or to an intrinsic perturbation power off in the same jet region. 
We speculate that, had this shutdown occurred in an even inner part of the jet, we would have not observed the outburst in the radio band, i.e.\ the event would have been even harder.

\section{Cross-correlation analysis}

In the previous section we saw that the brightness increase and consequent outburst
observed in the optical and near-IR bands were soon followed by similar events at mm and cm wavelengths.
We can of course estimate time delays by calculating the time separation between peaks in the different light curves.
But this procedure neglects the light curves behaviour as a whole, and strongly depends on the time sampling.
For instance, when comparing the $R$-band to the 1 mm light curve, 
we find a lag of about 1 month if we consider the December 2006 optical flare,
while the lag is reduced to a couple of weeks if we consider the outburst peak.
Hence, in order to obtain a better estimate of the average time delays between the flux changes observed at different frequencies, we apply the discrete correlation function (DCF; \citealt{ede88,huf92}) analysis.
Since we are analysing a limited time period with long data gaps, 
we use the cubic spline interpolations through the 
binned light curves that had already been shown in Fig.\ \ref{radop}.
In this way we certainly oversmooth the variations hidden by the solar conjunction period, 
but better define the major signals we are looking for. 
 
\begin{figure}
   \resizebox{\hsize}{!}{\includegraphics{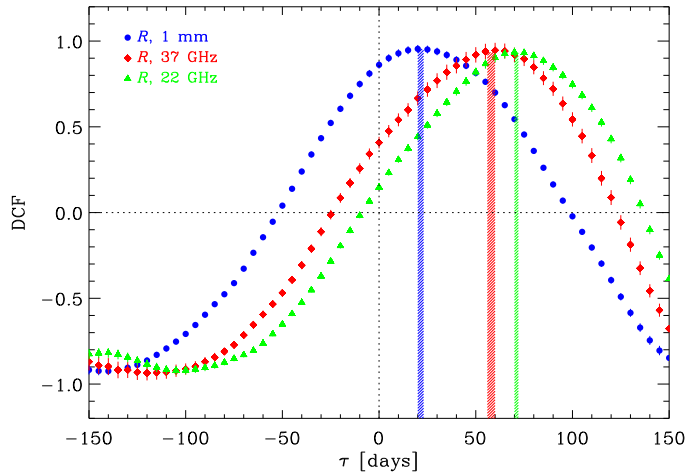}}
      \caption{Discrete correlation functions between the cubic spline interpolations through
the binned $R$-band and 1 mm light curves (blue dots), the $R$-band and 37 GHz ones (red diamonds), and the $R$-band and 22 GHz ones (green triangles).
Vertical strips highlight the range of time lags $\tau$ included between the location of the peak 
and the centroid of the distributions. }
         \label{ritardi}
   \end{figure}

Cross-correlation between the $R$-band and 1 mm, 37 GHz, and 22 GHz splines yields the DCF curves displayed in Fig.\ \ref{ritardi}.
In all cases the distribution of points shows a significant maximum ($\sim 1$)
at positive time lags $\tau$, which indicates a strong correlation with variations 
at the lower frequency lagging behind the higher-frequency ones.
The delays corresponding to the DCF peaks are 20, 60, and 70 days for the 1 mm, 37 GHz, and 22 GHz curves, respectively.
We obtain a better estimate of the time lag by calculating the centroid of the DCF \citep{pet98}, 
which is defined as $\tau_{\rm c}=\sum \tau_i \, {\rm DCF}_i  / \sum {\rm DCF}_i$ 
for all the points $i$ with ${\rm DCF}_i > k \, {\rm DCF_{peak}}$.
By considering $0.7 \le k \le 0.8$, we obtain lags between 22 and 23 days for the $R$-mm, 56--58 days for the $R$-37 GHz, and 70--72 days for the $R$-22 GHz cross-correlations.

However, we have already discussed in the previous section that the 22 GHz outburst seems to be affected by a sort of shutdown causing the sudden flux decrease. In the absence of this effect, the outburst peak and following decreasing phase would likely have been delayed. Hence, the 70--72 day delay indicated by the DCF appears as a ``compromise" between the more delayed rising phase and the less delayed decaying phase.
Indeed, by looking at the fourth panel of Fig.\ \ref{radop}, we see that the flux density increase at 22 GHz followed that at 37 GHz after about 25 days, the peak was reached 1 week later, and no lag is visible during the flux fading.

In the 14.5, 8, and 5 GHz light curves, the time of the outburst peak cannot be defined because of the inferior sampling, but we nonetheless infer that the ascending part of the outburst was delayed by $\sim 50$, $\sim 65$, and $\ga 90$ days with respect to the rising part of the 37 GHz outburst, while, again, no significant delay is visible in the dimming phase.

The application of the DCF method to auto-correlate the composite $R$-band historical light curve\footnote{This composite light curve has been obtained by considering the $R$-band data back to JD = 2447000, and $B$-band data converted into $R$-band ones (through the mean colour index $B-R=1.7$) before
that date \citep{rai06b}.} results in a reinforcement of the $\sim 8.5$ year signal already found by \citet{rai06b}.

\section{Spectral energy distributions}

One of the major issues raised by the works of \citet{jun04} and \citet{rai05,rai06b,rai06a} 
is the existence of an extra emission component in the SED of AO 0235+164, 
in addition to the synchrotron and inverse-Compton components.
This component is represented by a bump in the UV and soft X-ray energy range, and its nature is not yet clear.
In the case of quasar-type blazars, a similar feature is clearly visible only during faint states, and it is generally ascribed to thermal emission from an accretion disc. 
However, when these sources undergo
a flaring state, the bump tends to disappear, since it is overwhelmed by the synchrotron emission
\citep[see e.g.][for the case of 3C 454.3]{rai07b}.
In contrast, by comparing SEDs of AO 0235+164 obtained at various observing epochs in which X-ray observations were performed,
\citet{rai06b} pointed out that the presence of the extra component can also be inferred in bright states, and that the component appears to be variable. 
But they could not discriminate between the two most obvious interpretations, i.e.\
thermal emission from an accretion disc and an additional synchrotron component from an inner jet region.

With the aim of shedding some light on this, we selected epochs in which the source was in different brightness states and in which contemporaneous near-IR-to-UV data were available.
Magnitudes were corrected for both Galactic and foreground-galaxy absorption by following the 
prescriptions of \citet{jun04} (see Table 5 by \citealt{rai05}). 
We then converted de-reddened magnitudes into fluxes using the calibration by \citet{bes98} 
in the optical, and the zero-mag fluxes for the UVOT ultraviolet bands mentioned in Sect.\ 2.3.

\begin{figure}
   \resizebox{\hsize}{!}{\includegraphics{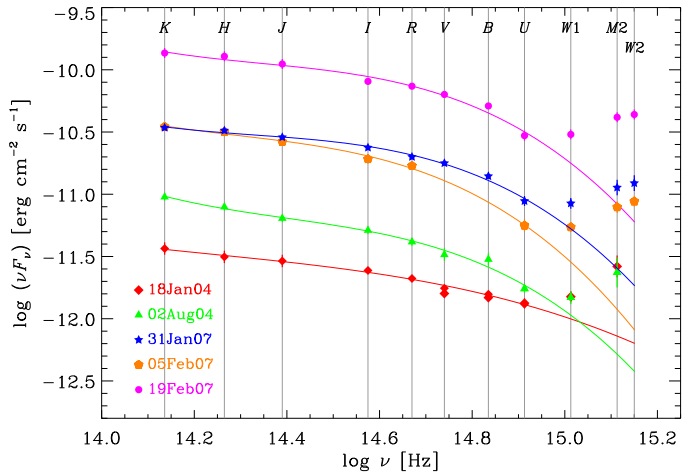}}
      \caption{Near-IR-to-UV spectral energy distributions at various epochs: 
the SED corresponding to the outburst maximum (February 19, 2007) is compared to 
the faint-state SEDs obtained during the January and August 2004 XMM-Newton pointings \citep{rai06b} 
and to two intermediate-state SEDs preceding the outburst peak.}
         \label{sed}
   \end{figure}

We show three of these SEDs in Fig.\ \ref{sed}, compared with those corresponding
to the XMM-Newton pointings of January and August 2004, published in \citet{rai06b}.
We built the brightest SED with data taken on February 19, 2007, at the culmination of the outburst; since there were no simultaneous $JHK$ data, we inferred near-IR magnitudes 
by considering the $R - JHK$ colour indices corresponding to the brightest near-IR level in
the light curves (see Fig.\ \ref{ottico-nir}).
In the other two cases (SEDs of January 31 and February 5, 2007) all data are contemporaneous (taken within few hours).
They show intermediate flux levels, with a noticeable spectral-slope variation:
the near-IR fluxes are nearly the same, but toward the UV the two SEDs diverge.

We found that the least-order polynomial that yields a good fit to the near-IR-to-optical part of the SEDs
is a cubic; we assume that these fits, which are displayed in Fig.\ \ref{sed}, 
can reasonably represent the contribution of the synchrotron component. 
The UV excess is clearly recognizable in all SEDs, and 
represents the extra emission component.
We estimated the contribution of this component in the UV bands as the difference between the total and the extrapolated synchrotron flux densities.

\begin{figure}
   \resizebox{\hsize}{!}{\includegraphics{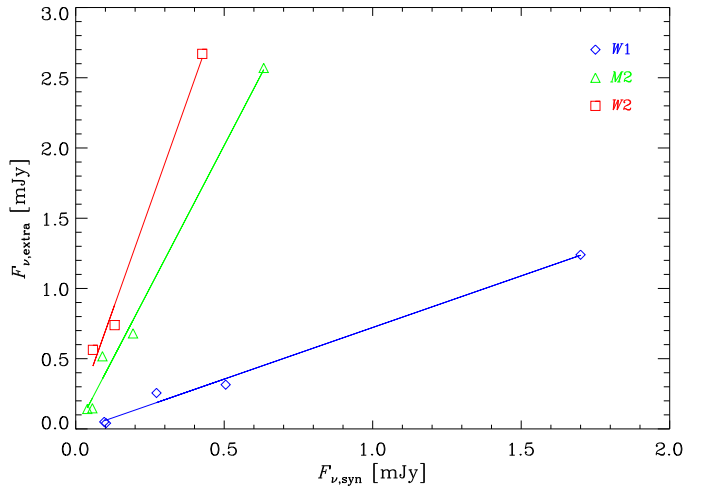}}
      \caption{The contribution to the flux density from the extra component
versus that from the synchrotron component in the three UVOT bands 
(see text for details). Lines represent linear fits to the data.}
         \label{extra}
   \end{figure}

Figure \ref{extra} shows the relationship between the extra-component and the synchrotron-component flux
densities for the three UV bands: a fair linear correlation can be seen (correlation coefficient $r=0.996, 0.995, 0.994$ for $W1$, $M2$, $W2$, respectively).
Notwithstanding the uncertainties affecting our fitting procedure, we think that this result indicates
that this extra component can hardly be explained in terms of thermal emission from the disc, which is
not expected to depend on the synchrotron radiation, but more likely it is produced by the same mechanism
that is responsible for the synchrotron emission.

\section{Discussion and conclusions}

The claim by \citet{rai01} of a possible quasi-periodic occurrence of the major radio and optical outbursts of AO 0235+164 every $5.7 \pm 0.5$ years led to a mobilization of the international blazar community to observe the next event, predicted to peak in early 2004. A WEBT campaign was organised, with a huge international participation \citep{rai05,rai06b,rai06a,rai07a}, but the source remained in a faint state and time series analysis on the updated light curves suggested a possible longer period of about 8--8.5 years, delaying the occurrence of the next outburst toward the 2006--2007 observing season.
An increased radio activity was registered in 2005--2006 \citep{bac07}, and a major optical outburst was finally observed in late 2006 -- early 2007, about 8.5 years after the previous major outburst peak, thus confirming the \citet{rai06b} prediction.

We observed the same event simultaneously in the near-IR and UV frequency ranges, and then at millimetric and centimetric wavelengths, with a progressive time delay toward lower frequencies.
We estimated that the 1 mm and 37 GHz outbursts lagged behind the $R$-band one by about 3 weeks and 2 months, respectively. 
This latter lag appears a factor $\sim 2$ 
longer than previously estimated \citep{rai05,rai06b} when considering the historical light curves until spring 2005.
This may reflect some real change either in the jet structure at still unresolved scales or in its energetics with respect to the past.
Indeed, there are sources that exhibit a characteristic behavior for many years and then suddenly change \citep[see e.g.][]{smi96,all96}.
Another possibility is that lack of data for long time intervals, chiefly in the optical bands because of solar conjunctions, affects the results of the DCF analysis.
Even in the case of the 1997--1998 outburst, which was intensively observed thanks to the
efforts of the just-born WEBT \citep{rai01}, the central part of the event was missed in the optical, and the dimming phase was poorly sampled at 37 GHz. As for the 2006--2007 outburst, notwithstanding the exceptional sampling, we cannot 
rule out that solar conjunction hid a further optical peak. This would shorten the radio lags.

In any case, the delays estimated in the present paper appear rather short if compared, 
for example, with the approximately twice-longer lags estimated by \citet{vil07} for the quasar-type blazar 3C 454.3 in correspondence of its 2004--2006 exceptional outburst, and by \citet{vil04b} and \citet{bac06} for BL Lacertae.
The short time delays in AO 0235+164 had already been noticed in the past \citep{web00,rai01}, and this was one of the reasons why microlensing was proposed as a possible explanation of the major variability events in this source \citep[see e.g.][and references therein]{rai07a}. 
In contrast, when interpreting the multifrequency variability of AO 0235+164 in the framework of the helical jet model by \citet{vil99}, the short time delays imply that the corresponding jet-emitting regions are only slightly misaligned  \citep{ost04}. If, besides the geometrical effect, the outburst is also produced by energetic processes inside the jet, short time lags imply that the emitting regions are also closeby.

At lower ($< 37$ GHz) radio frequencies, the outburst behaviour changes: while the rising phase is progressively delayed, as expected, the decaying phase is observed simultaneously at all wavelengths. In particular, the 22 GHz flux density reaches a maximum value similar to the 37 GHz flux density almost at the same time and then suddenly decreases. 
Although the observed general multifrequency properties of the outburst can be explained by models dealing with shocks propagating along an inhomogeneous jet \citep[see e.g.][]{hug89,val92}, the emission behaviour during the dimming phase might suggest that the scenario is more complicated than what is depicted in these models.
Indeed, it seems as if a kind of shutdown occurred in the jet between the mm emission region and the 37--22 GHz region. 
We can speculate that either a jet bending or an intrinsic power off of the perturbation may produce this effect, which
could also account for the different flavours of events observed in the historical light curves of the source: the harder outbursts, which are stronger at the higher frequencies, 
suggest that a shutdown (of any origin) has occurred between the higher- and the lower-frequency emitting regions.
Understanding whether the geometric or intrinsic scenario would be more plausible is not easy.
One can envisage that if the flux fading is due to a jet bending, the perturbation could continue to travel along the jet, following a misaligned path, until it could eventually be observed at radio wavelengths as soon as the jet (helical) path turns again toward the line of sight. For those sources, for which the Very Long Baseline Interferometry (VLBI) resolution can separate these regions, we should be able to observe a brightening of a VLBI knot.

We notice that our analysis of the multifrequency light curves in terms of harder/softer events presents some similarities with the generalized shock model by \citet{val92}, who distinguish between high- and low-peaking flares. 
We finally mention that \citet{hag07a} interpreted the photometric and polarimetric variations of AO 0235+164 during the December 2006 flare as due to the propagation of a transverse shock, accompanied by a small change in the viewing angle of the jet.

Another major issue we investigated is the existence of an extra emission component mostly contributing in the UV and soft X-ray energy range of the source SED. 
Swift-UVOT observations during the culminating phase of the optical outburst show that the behaviour of the UV emission strictly follows that of the optical one. Moreover, when constructing SEDs with simultaneous near-IR-to-UV data, and separating the synchrotron and extra-component 
contributions, we found that they are correlated. 
Although this result is affected by uncertainties due to the decomposition procedure, it nevertheless
suggests that an interpretation of the extra component in terms of radiation from the accretion disc is rather unlikely \citep[see][for the various possible interpretations]{rai06b,rai06a}. Also the hypothesis of two independent synchrotron components  appears now inadequate.
A still viable explanation could be that of anomalous absorption of the optical to near-UV emission in a spectrum that would otherwise be power-law-like from the near-IR to UV.
Indeed, examples of noticeable intrinsic absorption are found when analysing the spectra of broad absorption line (BAL) quasars, where several lines between $\sim$ 1000 and 1500 \AA\ (rest frame) can heavily reduce the flux in the corresponding spectral regions \citep[see e.g.][]{tur88}.
A further possibility is that the extra component is the result of inverse-Compton scattering of radio photons off the relativistic electrons producing the IR--optical emission. Indeed, as stated above, the close correlation and short time delays between the optical and radio variations suggest that a lot of radio photons are available in the optical emitting region.

\begin{acknowledgements}
We are grateful to Philip Hughes for useful discussions.
We acknowledge the use of public data from the Swift data archive.
This research has made use of data obtained through the High Energy Astrophysics Science Archive Research Center Online Service, 
provided by the NASA/Goddard Space Flight Center.
This work is partly based on observations made with the Nordic Optical Telescope, operated
on the island of La Palma jointly by Denmark, Finland, Iceland,
Norway, and Sweden, in the Spanish Observatorio del Roque de los
Muchachos of the Instituto de Astrofisica de Canarias.
This research has made use of data from the University of Michigan Radio Astronomy Observatory,
which is supported by the National Science Foundation and by funds from the University of Michigan.
The Mets\"ahovi team acknowledges the support from the Academy of Finland.
AZT-24 observations are made within an agreement between  Pulkovo,
Rome and Teramo observatories.
The St.-Petersburg team was supported by the Russian Foundation for Basic Research, grant
08-02-00245.
The Torino team acknowledges financial support by the Italian Space Agency through contract 
ASI-INAF I/088/06/0 for the Study of High-Energy Astrophysics. 
University of Joensuu acknowledges co-operation with local Amateur Astronomer Association Seulaset.
ACG work is supported by NNSF of China grant No.10533050.
JHF's work is supported by the National Natural Science Foundation of China (10573005,10633010) and the 973 project (No. 2007CB815405).
\end{acknowledgements}

\end{document}